\documentclass[proof]{WileyASNA-v1}

\articletype{Article Type}

\received{27 January 2020}
\revised{}
\accepted{}

\begin{document}

\title{Modelling and simulations of supernova remnants: a short review focused on recent progress in morphological studies}

\author[1,2]{Gilles Ferrand*}

\address[1]{\orgdiv{Astrophysical Big Bang Laboratory (ABBL)}, \orgname{RIKEN}, \orgaddress{\state{Saitama}, \country{Japan}}}

\address[2]{\orgdiv{Interdisciplinary Theoretical and Mathematical Sciences Program (iTHEMS)}, \orgname{RIKEN}, \orgaddress{\state{Saitama}, \country{Japan}}}

\corres{*\email{gilles.ferrand@riken.jp}}

\abstract{Supernova remnants (SNRs) are the outcome of supernovae 
(SNe, either core-collapse or thermonuclear). The remnant results from the interaction
between the stellar ejecta and the ambient medium around the progenitor
star. Young SNRs are characterized by strong shocks that heat and
ionize the gas, generate magneto-hydrodynamic turbulence, and accelerate
particles to relativistic energies. They radiate at all wavelengths,
especially in the X-ray domain, where spectro-imaging observations
can provide a wealth of information. 

This paper presents recent progress in the modelling of SNRs, 
particularly by the means of numerical simulations, 
and with a focus on three-dimensional aspects. 
In the first part we will consider SNRs as producers of cosmic rays (CRs).
If SNRs are accelerators efficient enough to power the Galactic component
of CRs, this must have a visible impact on their dynamics, and therefore
on the thermal emission from the plasma, as well as on their non-thermal
emission. In the second part we will consider SNRs as probes of the
explosion mechanism. The time has come to connect multi-dimensional simulations of SNe
and simulations of SNRs, opening the possibility to study the explosion
mechanism via the dynamics and morphology of SNRs.}


\maketitle

\section{Introduction to SNRs}\label{sec:intro}

A~supernova (SN) marks the explosion of either a white dwarf (Type Ia SNe) or a massive star (Type II and Ib/c SNe). 
Supernova remnants (SNRs) are produced from the interaction between stellar ejecta and the circumstellar medium (CSM) and interstellar medium (ISM). As such, SNRs are a key link between the lifecycle of stars and the cycle of matter in the galaxy. The ejecta, enriched in heavy elements produced during the stellar life and during the explosion itself, expand over tens of parsecs over tens of thousands of years, until the SNR dissolves in the ISM. Their initially highly supersonic motion triggers a powerful blast wave, which heats and ionizes the matter at the same time as it accelerates a fraction of particles to relativistic energies. For a review of SNR properties we refer the reader to \cite{Reynolds2008a} and \cite{Vink2012a}, and references therein. We summarize the main points here.

From multi-wavelength observations, especially in radio and X-rays, SNRs can be classified in different types. In the ``shell'' type SNRs, like Tycho's SNR in our Galaxy or SNR 0509-67.5 in the LMC, we observe the shell of shocked matter at the edge of the remnant. In so-called ``filled-centre'' SNRs, or ``plerions", we mostly see the non-thermal emission from a pulsar wind nebula (PWN), the prototype being the Crab nebula; a pulsar can also generate a bow-shock nebula (like the Guitar nebula). A~``plerionic composite'' is basically a PWN inside a shell, the prototypical example being SNR G21.5-0.9. There are other kinds of composites though: so called ``mixed morphology'' SNRs like W49B are centrally peaked, but their emission is thermal; some SNRs qualify as both thermal and non-thermal composites (e.g. IC~443). In this short review we focus on the canonical shell-type remnants.

SNRs evolve over different phases. A~simplified picture is a follows. Initially their dynamics is ejecta dominated, followed by a universal self-similar solution known as the Sedov-Taylor phase. These phases are said to be non radiative, in the sense that radiation, although detectable, does not affect the dynamics. After the plasma has cooled down enough the radiative losses become important -- if the SN occurs in a dense environment this phase may actually happen early on. In late phases the SNR is a pressure-driven bubble bounded by a thin shell, that eventually is only momentum-driven, and dissolves in the ISM. For particle acceleration to be efficient one needs strong shocks, so young SNRs; and to make the link with SN explosion one wants to catch them as early as possible, so our focus here will be on young SNRs.

The structure of a young SNR shell is delineated by two shocks: the forward shock travelling in the CSM/ISM, and the reverse shock travelling backward in the ejecta due to their deceleration. The interface between the shocked ejecta and the shocked CSM/ISM, called the contact discontinuity, is hydrodynamically unstable: the Rayleigh-Taylor instability (RTI) produces a distinctive pattern of fingers and holes \citep{Chevalier1992a}. It is thus necessary to make a multi-dimensional modelling if one wants to reproduce the morphology of the SNR, even assuming simplified spherical initial conditions.

A~young SNR is a broad-band emitter. Chiefly, the hot plasma is shining in X-rays, the shock may also be traced in the optical (H$\alpha$ line).\footnote{The shocked ejecta can also detected in the optical: see the discovery by \cite{Seitenzahl2019}.} Energetic particles radiate at both ends of the spectrum: electrons produce synchrotron radiation in a magnetic field, from radio to gamma energies, and inverse Compton radiation on the ambient photon fields, in $\gamma$-rays; protons induce pion decay from collisions with ambient matter, which also produces $\gamma$-ray photons. See \cite{Reynolds2008a} and \cite{Vink2012a} for reviews of observations from the high-energy perspective, and \cite{Lopez2018} for a review focused on morphology and kinematics.
\footnote{A~catalogue of high-energy observations of SNRs, now including images in radio and X-ray bands, is available at \url{http://snrcat.physics.umanitoba.ca}.}

In this paper we outline important processes that shape young SNRs, intrinsic to the SNR phase, or coming from the SN phase. We illustrate, using Tycho as a template, how a multi-dimensional modelling can shed light on the observed morphology of nearby young SNRs.

\section{SNRs as particle accelerators}\label{sec:accel}

SNRs are commonly believed to be the main producers of cosmic rays (CRs) in the Galaxy \citep{Blasi2013c}. The original argument was the energy budget: it requires channeling only of the order of 10\% of SNe energies into particles. However to this day there is little evidence that SNR shocks actually accelerate particles up to the so-called ``knee'' observed in the CR spectrum around the PeV, which is considered to be a minimum requirement for Galactic sources. We do know that particles can be accelerated at collisionless magnetized astrophysical shocks, via the process of diffusive shock acceleration (DSA), which naturally produces power-law spectra as observed. However the actual shape of the spectrum will be more complicated when acceleration is efficient \citep{Malkov2001c}. We routinely observe energetic electrons (in radio, and in X-rays), but these make a tiny fraction of CRs. Since the coming of age of TeV astronomy (with instruments like H.E.S.S., MAGIC, VERITAS, and HAWC) we can now also observe energetic protons, but it remains challenging to disentangle the emissions from leptons and hadrons. 
Now, if CRs are efficiently accelerated by the blast wave, it must impact its dynamics: the fluid becomes more compressible, as particles become relativistic and energy leaks from the system. We thus have to solve the evolution of a non-linearly coupled system \citep{Decourchelle2000a}. It follows that CRs are not merely a by-product, but an integral ingredient of SNRs.

\subsection{Coupling hydro evolution and particle acceleration}\label{sec:accel-coupling}

DSA is a complex process which involves: the shock wave in the plasma assumed to follow a thermal distribution, energetic particles (to become CRs) that constitute a non-thermal tail in the distribution, and magnetic waves that couple the two populations. Modelling DSA requires numerical simulations with different levels of treatment for different purposes \citep{Marcowith2020}. The deepest level, particle-in-cell (PIC) simulations, despite recent progress, is still impractical at astrophysical scales \citep{Vladimirov2008a,Ellison2011a}. Commonly used approaches (see a comparison in \cite{Caprioli2010c}) couple hydrodynamic and kinetic treatments, with a fluid description for the shock (conservation laws for mass, momentum, energy) and a kinetic description for the particles (advection-diffusion equation for their distribution function $f(x,p,t)$). The time-dependent evolution of DSA has been studied in 1D, but for 3D simulations of SNRs semi-analytical models have been used instead for particle acceleration, chiefly the one from Blasi et al (\citeyear{Blasi2004a,Blasi2005a}). 

The main effect of efficient particle acceleration on SNR dynamics is that the shocked region gets more compact, meaning both smaller (as observed in young Galactic SNRs) and denser (which may be more difficult to constrain observationally). This was investigated in 3D in \cite{Blondin2001a}, by varying the adiabatic index of the fluid to mimic the presence of relativistic particles and their escape. This was studied in 1D (so without the RTI) in \citep{Ellison2007a}, using an acceleration model to predict the pressure of energetic particles at the shock, from which an effective adiabatic index for the fluid is derived as a function of time. The two approaches were combined in \cite{Ferrand2010a}: the SNR evolution is computed in 3D up to an age of 500~yr with the hydro code RAMSES, in a comoving grid in order to factor out the global SNR expansion; at each time step Blasi's acceleration model predicts the spectra of particles at the shock, from which the back-reaction is imposed. Resolving both the RTI and the DSA makes it possible to generate the synthetic SNR images presented in the next section.

\subsection{Computing the emission from the SNR}\label{sec:accel-emission}

\begin{figure}[ht]
\centerline{\includegraphics[width=0.85\columnwidth]{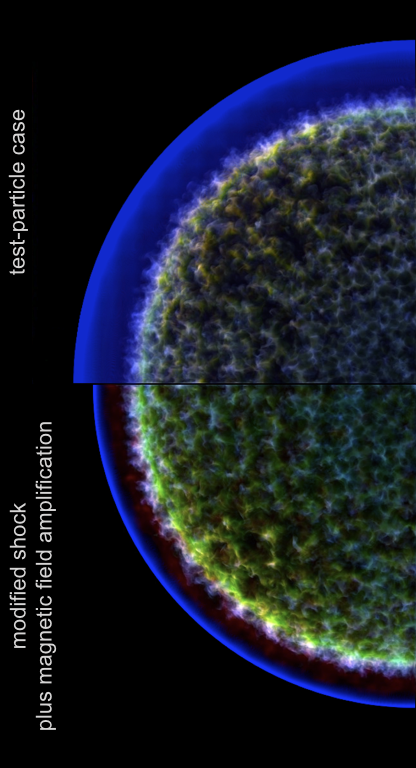}}
\caption{Simulated X-ray maps for a Tycho-like SNR at 500~yr. The greenish fluffy central part is the thermal emission from the shocked ejecta (RGB composite in three energy bands). The outer blue layer is the non-thermal emission from the relativistic electrons. The two sides compare the cases without (at the top) and with (at the bottom) taking into account the back-reaction of accelerated particles on the shock dynamics and on the magnetic field. 
\label{fig:accel}}
\end{figure}

Now we briefly list the points that need to be considered in order to be able to compute the broadband emission from a (young, CR-accelerating) SNR. We illustrate this with the simulations from \cite{Ferrand2012g} for the thermal emission and \cite{Ferrand2014b} for the non-thermal emission. A~composite of mock SNR maps from these two papers, in the X-ray band, is shown in Figure~\ref{fig:accel}.\footnote{For the X-ray image of Tycho's SNR, the reader is referred to the paper by \cite{Warren2005a} and the Chandra website at \url{http://chandra.harvard.edu/photo/2005/tycho/}.} We refer the reader to the references in these papers for further details.

The thermal emission depends on the plasma density (squared), on the electron temperature, and on the ionization state of all the species present in the plasma. Regarding the electron temperature a difficulty is that what simulations readily produce are the proton (or ions) temperatures, an assumption is necessary for electron heating at the shock front \citep{Ghavamian2007a}, then it is expected that progressive equilibration will happen behind the shock via Coulomb interactions. Regarding ionization, it is important to note that the plasma in young SNRs is in a non-equilibrium ionization (NEI) state. In principle one has to solve a time-dependent system of coupled equations, this can be done in 1D (e.g. \cite{Patnaude2009a, Patnaude2010a}), this is impractical in 3D due to memory requirements, but one can instead use the exponentiation method \citep{Smith2010a}, in post-processing. Knowing the plasma state, one can then use an emission code, such as AtomDB/APEC, SPEX, or Chianti, in order to compute the emission from the plasma: thermal Bremsstrahlung and emission lines. Looking at the (spatially integrated) X-ray spectrum from the SNR, the main effect of particle back-reaction at the forward shock is to reduce the emission of the plasma, all the more so at the highest energies. This enhances the visibility of the emission lines from heavy elements in the ejecta.

The non-thermal emission depends on: the plasma density, to compute pion decay; the magnetic field, to compute the synchrotron radiation, and the CMB or other ambient photon fields, to compute the inverse Compton emission. For the magnetic field, a MHD description is appropriate downstream of the shock, where the field can be assumed to be frozen in the plasma, whereas a kinetic description is in order at the shock front, where the field is believed to be amplified as part of the acceleration process itself \citep{Schure2012b}. One also needs the particle spectra (both electrons and protons), which are computed by the acceleration model at the shock front, and have to be transported downstream, taking into account adiabatic losses and radiative losses. Looking at the broadband spectrum from the SNR, the back-reaction of particles affects the maximum energy reached in $\gamma$-rays (not so much in X-rays). Looking at maps, it alters the spatial distribution at different wavelengths: radio emission, that probes low-energy electrons, is unaffected; X-ray emission, that probes the electron cut-off, gets concentrated close to the shock front (compare the two sides of Figure~\ref{fig:accel}), as observed in several young Galactic SNRs (see \citealt{Warren2005a} for the case of Tycho); $\gamma$-ray emission is even narrower, but this cannot be resolved by current instruments. 

\subsection{Main take-away points}\label{sec:accel-points}

In the context of probing particle acceleration in young SNRs, modelling work has shown that energetic protons, even though they do not radiate as efficiently as electrons, 1/ impact the dynamics of the shock wave via their pressure, and therefore the thermal emission from the shell of shocked matter (in the X-rays, also in the optical); and 2/ impact the evolution of the magnetic field, and therefore the non-thermal emission from the electrons (in radio to X-rays and $\gamma$-rays).

\section{SNRs as probes of the explosion}\label{sec:explosion}

The two main types of SNe correspond to different explosion mechanisms, each with competing theoretical models. 
For Type Ia the consensus is the thermonuclear explosion of a white dwarf, but this can happen in a number of different ways, and involve different kinds of companion stars (see \citealt{Hillebrandt2013a} for a review). Type Ia SNe are important since they serve as ``standard candles'' for cosmology, yet to this date there is still no consensus on the explosion mechanism(s).
For Type II and Ib/c, the collapse of the core of a massive star is invoked, a major problem being to revive the shock, which was observed to get stalled inside the dying star. A~promising candidate for this is neutrinos, but the mechanisms is still not well understood. An important aspect in recent numerical simulations has been the increase in dimensionality (see \citealt{Janka2016} for a review). In a realistic 3D setup some, but not all, models are found to explode. Successful explosions have a complex structure, does it impact the morphology of the remnant? An interesting question is what can the observed morphology of the young SNR tell us about the explosion mechanism. A~number of studies have been done linking the SN to the SNR, relying on a 1D modelling (see e.g. the grid of models in \cite{Badenes2005b}, and the end-to-end simulations in \cite{Patnaude2017}). Using the radial abundance profiles from the SN models, such models can address the global properties of the SNR emission. Being computationally affordable, they are used to explore the parameter space. We present in this section recent works that are specifically bridging multi-dimensional SN studies and SNR studies.

\subsection{Core-collapse case}\label{sec:explosion-CC}

Pioneering studies in this field have been conducted by \cite{Orlando2015a} and \cite{Orlando2016a}, aimed respectively at explaining SN~87A and Cas~A.

For the case of Cas~A, \cite{Orlando2016a} mapped a 1D SN model to 3D, added some asymmetries by hand, evolved it to hundreds of years, and computed the X-ray emission. The initial conditions were tweaked until they could reproduce the overall morphology of the SNR (see their Figure~11), which does show that the bulk of the asymmetries observed can be intrinsic to the explosion. Can we get such distorted shapes from core-collapse SN explosions? Certainly, see as an example the grid of parametrized neutrino-driven models from \cite{Wongwathanarat2015a}, evolved until the shock breakout (see their Figure~7). Taking the W15-2 model, stripping it of its H envelope to make it closer to a type IIb, and further excising the central part (which cannot have been shocked yet and so is not visible in X-rays), \cite{Wongwathanarat2017a} obtained three blobs of 56Ni, similar to the Fe blobs observed in Cas~A (Fe is mostly the product of the decay of 56Ni). This (chance) result is much encouraging. That being said, the images in the paper are only density maps, and at around one day (see their Figure~8). The work remains to be done, to evolve the SNR over a few hundred years and to compute its X-ray emission, as was done in the Orlando et al papers and outlined in the previous section. The expertise on hot plasmas of the X-ray community is critical for making the SN-SNR connection.

For the case of SN~87A, \cite{Orlando2015a} also relied on a 1D SN model, and performed 3D simulations of the SNR phase, including the pre-supernova nebula (clumpy ring and HII region). They were able to reproduce the main observables, to disentangle the different contributions to the X-ray emission, and to predict its future evolution. This work has just been updated: \cite{Ono2020} conducted 3D simulations of the matter mixing in the SN phase, and \cite{Orlando2020} followed the evolution into the SNR phase. They provide compelling evidence in favour of the merger scenario, based on observational constrains on 3D anisotropies of the explosion at very different times.

\subsection{Thermonuclear case}\label{sec:explosion-TN}

\begin{figure}[t]
\centerline{\includegraphics[width=0.9\columnwidth]{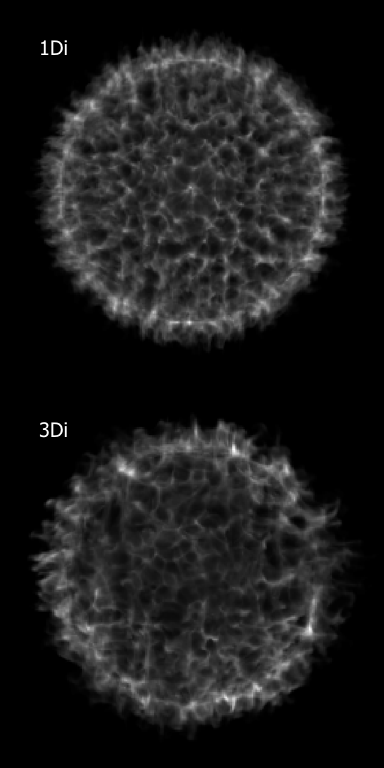}}
\caption{Mock images for a SNR at an age of 500 yr (the quantity displayed here is the density squared of the shocked ejecta, which is a proxy for the broad-band X-ray emission). The two images compare the case of initial conditions that are effectively 1D (top) versus fully 3D (bottom), as obtained from the N100 SN explosion model.
\label{fig:explosion}}
\end{figure}

We know that thermonuclear SNRs tend to be more symmetric than core-collapse SNRs -- this was quantified by Lopez (\citeyear{Lopez2009a,Lopez2011a}). However thermonuclear explosions are more complicated than traditionally assumed when studying the SNR phase. Simulating SNe and simulating SNRs have historically been separate concerns: SN people want to get an explosion, and stop their simulations after maybe 100 seconds; SNR people do not do the explosion per se, they start a few years after. The multi-dimensional SNR simulations cited in section~2 assumed idealized initial conditions, spherically symmetric, with a simple radial dependence (either a power-law profile, allowing the use of the semi-analytical model by \cite{Chevalier1983a}, or an exponential profile, investigated by \cite{Dwarkadas1998a}). A~natural idea is to use the output from a SN simulation as the input for a SNR simulation. In~\cite{Ferrand2019} we used the popular N100 model from \cite{Seitenzahl2013a}, comparing for initial conditions: the full 3D initial profiles, an angularly averaged version (effectively 1D), as well as the Chevalier model (1D). Note that during the SNR phase one always get the RTI growing, even from smooth profiles; in a more realistic scenario this happens on top of already irregular ejecta, as imposed by the 3D nature of the deflagration and detonation fronts. Overall the SNR structure regularizes itself over time -- eventually all models will converge to the Sedov-Taylor solution, which does not depend on the explosion details. We observed that the SN imprint is clearly visible at 100~yr, with some differences still visible at 500~yr. To quantify this, we extracted the three wavefronts: forward shock, contact discontinuity, reverse shock, projected their relative radial variations on the sphere, and expanded them in spherical harmonics to compute the angular power spectrum. For the contact discontinuity, the two components can be separated: the RTI, rising over time from the smallest scales to larger scales (for both 3D and 1D initial conditions), and the SN modes, at large scales and decaying. At an age of 500~yr the angular spectrum happens to look like a single distribution, as if the result of enhanced RTI, but really the large scale modes are echoes of the SN explosion. Figure~\ref{fig:explosion} shows a first estimate for the X-ray emission: projection along the line of sight of the density squared. We still have to compute the actual thermal emission in different energy bands taking into account the nucleosynthesis products (work in progress), but Tycho's SNR looks more like the map made from the realistic SN model.

In the future this technique can be used to make comparisons between different SN explosion models. The N100 model is a delayed detonation, \cite{Seitenzahl2013a} actually presented a grid of such models, with the ignition setup parametrized, which produces different initial asymmetries and nucleosynthesis yields. There are other kinds of models for the explosion of a single white dwarf, like pure deflagrations or pure detonations. And yet others when considering the double-degenerate scenario, for example the double-detonation model of \cite{Tanikawa2018}, where one can see the imprint of the companion just after the explosion.

\subsection{X-ray image analysis}\label{sec:explosion}

We now discuss how one may distinguish the different models from the X-ray images. (Alas we cannot presently do a full 3D analysis as in the simulations, because most of the information we have is in the plane of the sky.)
Common techniques for image analysis are based on expanding the surface brightness on some nice basis of functions, as done in Fourier analysis, and in the multipole expansion of \cite{Lopez2009b} (``power ratio method'').
 We are currently exploring such techniques in our group (work by H.~Iwasaki).
This category also includes wavelet functions (also investigated in the Lopez et al paper), an extension of which is used for the recently developed technique of ``general morphological component analysis'' \citep{Picquenot2019}. 
Also techniques from topology are being investigated: \cite{Sato2019} applied the ``genus statistics'' (technically, the Euler-Poincar{\'e} characteristic on the excursion set) to the SNR simulations presented in \cite{Williams2017}; they were able to mathematically quantify (the obvious) that Tycho's SNR is not of the smooth kind. Their SNR simulations from smooth versus clumpy initial profiles are actually looking very similar to the ones in \cite{Ferrand2019}, the difference being that their initial conditions were setup by hand.

Finally, we note that in the \cite{Williams2017} and \cite{Sato2017} observational papers, the primary aim was to reconstruct the 3D structure of the SNR from the velocity of selected knots, measured from the Doppler shift along the line of sight. This is a difficult task in X-rays using current instrumentation, but if one has the energy resolution of the (lost) Astro-H/Hitomi mission, and the sensitivity of the (planned) Athena mission, then, assuming one can de-blend the forest of lines that will be resolved, one can in principle reconstruct the velocity profile at the same level of detail as can be predicted from the 3D simulations (see an example for Tycho in \cite{Decourchelle2013a}, an Athena+ supporting paper). This will open a new dimension for the study of the explosion mechanism.

\section{Conclusion}

In conclusion observations of young SNRs, in particular in the X-ray domain, reveal a wealth of information on the original explosion mechanism and on the acceleration of particles at the blast wave, which can be used to constrain the origin of heavy elements and of cosmic rays in the Galaxy.

To make major progress, we need observations that reveal the 3D structure of SNRs, as was done for Cas~A \citep{DeLaney2010a, Milisavljevic2015a}. See \cite{Lopez2018} for the exciting prospects in the X-ray domain. This will realize the full potential of 3D simulations, from the SN to the SNR. Accordingly, we shall develop 3D visualization and analysis techniques.\footnote{e.g. using virtual reality, \cite{Ferrand2018}.}


\section*{Acknowledgments}

The author thanks the organizers of the XMM-Newton 2019 Science Workshop ``Astrophysics of Hot Plasma'' at ESAC for their invitation. 
The author acknowledges his many collaborators on this topic, including Anne Decourchelle, Jean Ballet, Samar Safi-Harb, Shigehiro Nagataki, Donald Warren, Masaomi Ono, Friedrich R{\"o}pke, Ivo Seitenzahl, Hiroyoshi Iwasaki, Yasunobu Uchiyama, Toshiki Sato.

This work was supported by the \fundingAgency{"Pioneering Program of RIKEN for Evolution of Matter in the Universe (r-EMU)"}.



\subsection*{Financial disclosure}

None reported.

\subsection*{Conflict of interest}

The authors declare no potential conflict of interests.



\bibliography{XMM_review}%

%

\end{document}